\documentclass[12pt]{article}
\usepackage{a4p}
\usepackage{cite}
\usepackage{epsfig}
\usepackage{amsmath} 

 \def\exeff1{\ensuremath{0.0736(.0018) 0.0105(.0007) 0.0043(.0005)}}
 \def\exeff2{\ensuremath{0.0081(.0009) 0.0360(.0019) 0.0019(.0004)}}
 \def\exeff3{\ensuremath{0.0028(.0006) 0.0008(.0003) 0.0353(.0021)}}
 \def\exbra{\ensuremath{(9.1\pm 0.9\pm 0.6})\times 10^{-3}}
 \def\exbrb{\ensuremath{(3.6\pm 1.3\pm 1.0})\times 10^{-3}}
 \def\exbrc{\ensuremath{(3.3\pm 0.9\pm 0.7})\times 10^{-3}}

 \def\exefii{\ensuremath{7.36\pm0.18}}
 \def\exefji{\ensuremath{1.05\pm0.07}}
 \def\exefki{\ensuremath{0.43\pm0.05}}
 \def\exefij{\ensuremath{0.81\pm0.09}}
 \def\exefjj{\ensuremath{3.60\pm0.19}}
 \def\exefkj{\ensuremath{0.19\pm0.04}}
 \def\exefik{\ensuremath{0.28\pm0.06}}
 \def\exefjk{\ensuremath{0.08\pm0.03}}
 \def\exefkk{\ensuremath{3.53\pm0.21}}
 \def\exmcii{0.24}
 \def\exmcjj{0.28}
 \def\exmckk{0.24}
 \def\exbiii{0.14}
 \def\exbijj{0.05}
 \def\exbikk{0.05}

 \def\expOii{0.14}
 \def\expOjj{0.27}

 \def\exmoii{0.00}
 \def\exmojj{0.39}
 \def\exmokk{0.17}

 \def\exdeii{0.21}
 \def\exdejj{0.11}
 \def\exdekk{0.33}
 \def\exbtii{0.29}
 \def\exbtjj{0.50}
 \def\exbtkk{0.31}
 \def\exklii{0.40}
 \def\exkljj{0.68}
 \def\exklkk{0.42}
 \def\exttii{0.62}
 \def\exttjj{1.02}
 \def\exttkk{0.68}

 \def\biex1{0.986\pm0.009}
 \def\biex2{0.995\pm0.015}
 \def\biex3{0.999\pm0.015}
 \def\nnkpi{178}
 \def\nkpi0{ 81}
 \def\nkkpi0{ 41}
 \def\nklong{305}
 \def\klinbg{24}
 \def\klinqq{6}
 \def\biexi{\mbox{$0.99\pm0.01$}}
 \def\biexj{\mbox{$1.00\pm0.01$}}
 \def\biexk{\mbox{$1.00\pm0.01$}}

 \def\brkoi{{(9.33\pm0.68\pm0.49})\times 10^{-3}}
 \def\brkoj{{(3.24\pm0.74\pm0.66})\times 10^{-3}}
 \def\brkok{{(3.30\pm0.55\pm0.39})\times 10^{-3}}
 \def\ntpr{100925}

 \def\nta {  85789}

 \def\idt{349}
 \def\adt{199}
 \def\bdt{ 67}
 \def\cdt{ 83}

 \def\abrz{( 9.6\pm 1.0\pm 0.7)\times 10^{-3}}
 \def\bbrz{( 3.0\pm 0.9\pm 0.9)\times 10^{-3}}
 \def\cbrz{( 3.3\pm 0.7\pm 0.5)\times 10^{-3}}

 \def\effaa{8.94\pm 0.18}
 \def\effba{0.48\pm 0.05}
 \def\effca{0.99\pm 0.07}
 \def\effab{2.88\pm 0.18}
 \def\effbb{5.32\pm 0.24}
 \def\effcb{0.93\pm 0.10}
 \def\effac{1.46\pm 0.13}
 \def\effbc{0.27\pm 0.06}
 \def\effcc{8.44\pm 0.29}

 \def\axa{0.10}
 \def\bxa{0.09}
 \def\cxa{0.07}
 \def\axb{0.23}
 \def\bxb{0.17}
 \def\cxb{0.15}
 \def\axc{0.40}
 \def\bxc{0.44}
 \def\cxc{0.21}
 \def\axd{0.46}
 \def\bxd{0.66}
 \def\cxd{0.18}
 \def\axe{0.26}
 \def\bxe{0.10}
 \def\cxe{0.32}
 \def\axf{0.22}
 \def\bxf{0.22}
 \def\cxf{0.00}
 \def\axg{0.00}
 \def\bxg{0.31}
 \def\cxg{0.16}
 \def\axt{0.74}
 \def\bxt{0.91}
 \def\cxt{0.48}

\def\afbias{0.99 \pm 0.01}
\def\bfbias{1.01 \pm 0.03}
\def\cfbias{0.99 \pm 0.02}

\newcommand{\Pem}{\ensuremath{\mathrm{e^-}}}
\newcommand{\Pep}{\ensuremath{\mathrm{e^+}}}
\newcommand{\Pgmm}{\ensuremath{\mathrm{\mu^-}}}
\newcommand{\Pgmp}{\ensuremath{\mathrm{\mu^+}}}
\newcommand{\PZz}{\ensuremath{\mathrm{Z^0}}}
\newcommand{\PKzL}{\ensuremath{\mathrm{K^0_L}}}
\newcommand{\PKzS}{\ensuremath{\mathrm{K^0_S}}}
\newcommand{\PKz}{\ensuremath{\mathrm{K^0}}}
\newcommand{\PK}{\ensuremath{\mathrm{K}}}
\newcommand{\PKm}{\ensuremath{\mathrm{K^-}}}
\newcommand{\PaKz}{\ensuremath{\mathrm{\overline{K}^0}}}
\def\nut{\nu_\tau}
\def\tm{\tau^-}
\def\tp{\tau^+}

\def\tt	{\tp\tm}
\def\ee {\Pep\Pem}
\def\mm {\Pgmp\Pgmm}
\def\qq {\mathrm{q}\bar{\mathrm{q}}}

\def\eeee {\ee \!\! \rightarrow \! \ee}
\def\eemm {\ee \!\! \rightarrow \! \mm}

\def\eeqq {\ee \!\! \rightarrow \! \qq}

\def\eeeeee {\ee \!\! \rightarrow \! (\ee)\ee}
\def\eeeemm {\ee \!\! \rightarrow \! (\ee)\mm}

\def\ebeam     {\mbox{$E_{\mathrm{beam}}$}}
\def\ehb       {\mbox{$E_{\mathrm{HB}}$}}

\def\dedx      {{\mathrm{d}} E/{\mathrm{d}}x}

\def\etal      {{\it et~al.}}
\def\ppion     {\mathrm{P}_\pi}
\def\pkaon     {\mathrm{P}_{\mathrm{K}}}
\def\shb       {S_{\mathrm{H}}}

\def\xm        {\mathrm{X}^-}
\def\kk        {\mathrm{K}}
\def\km        {\kk^-}
\def\kz        {\kk{}^0}
\def\ks        {\kk{}^0_{\mathrm S}}
\def\kl        {\kk{}^0_{\mathrm L}}
\def\kzb       {\overline{\kk}{}^0}
\def\ksb       {\overline{\kk}{}^0_{\mathrm S}}
\def\klb       {\overline{\kk}{}^0_{\mathrm L}}

\def\pkz       {\pi^- \kzb}
\def\pkzp      {\pi^- \kzb [\geq 1\pi^0]}
\def\kkz       {\kk^- \kz  [\geq 0\pi^0]}

\def\tauks     {\tm \rightarrow  \xm \; \ksb \nut } 
\def\taukl     {\tm \rightarrow  \xm \; \klb \nut } 
\def\taupks    {\tm \rightarrow  \pi^- \ksb \nut } 
 
\def\taupkl    {\tm \rightarrow  \pi^- \klb \nut }

\def\taupkzkz  {\tm \rightarrow  \pi^- \kzb  \kz   \nut } 
\def\taupkzkzp {\tm \rightarrow  \pi^- \kzb  \kz \pi^0 \nut } 
\def\taupkz    {\tm \rightarrow  \pi^- \kzb    \nut } 
\def\taupkp    {\tm \rightarrow  \pi^- \kzb  \pi^0  \nut } 
\def\taupkpp   {\tm \rightarrow  \pi^- \kzb \pi^0 \pi^0  \nut } 
\def\taupkzp   {\tm \rightarrow  \pi^- \kzb \mbox{\small $[ \geq 1\pi^0]$}  \nut } 
\def\taukkz    {\tm \rightarrow  \km \kz {\mbox{\small $[\geq 0\pi^0]$}} \nut } 
\def\taukka    {\tm \rightarrow  \km \kz \nut } 
\def\taukkb    {\tm \rightarrow  \km \kz {\mbox{\small $[\geq 1\pi^0]$}} \nut } 
\def\taukkc    {\tm \rightarrow  \km \kz \pi^0 \nut } 
\def\taukkd    {\tm \rightarrow  \km \kz \pi^0\pi^0 \nut } 

\def\taukstar  {\tm \rightarrow  \kk^*(892)^-  \nut }

\def\thpihz    {\ensuremath{\tau^{-}\rightarrow \mathrm{h}^{-} \pi^0\nu_\tau}}

\def\trho      {\ensuremath{\tau^{-}\rightarrow\rho^{-} \nu_\tau}}
\begin{document}

\begin{titlepage}
\begin{center}{\large   EUROPEAN LABORATORY FOR PARTICLE PHYSICS
}\end{center}\bigskip
\begin{flushright}
       CERN-EP/99-154   \\ 29 October 1999 \\ Journal Version \\ 22 November 1999
\end{flushright}
\bigskip\bigskip\bigskip\bigskip\bigskip
\begin{center}{\huge\bf  Tau Decays with Neutral Kaons
}\end{center}\bigskip\bigskip
\begin{center}{\LARGE The OPAL Collaboration
}\end{center}\bigskip\bigskip

\begin{abstract}

The branching ratio of the $\tau$ lepton to a neutral kaon meson
is measured from a sample
of approximately 200,000 $\tau$ decays recorded by the OPAL detector
at centre-of-mass energies near the $\mathrm{Z}^0$ resonance.
The measurement is based on two samples which identify
one-prong $\tau$ decays with \PKzL\ and \PKzS\ mesons.
The combined branching ratios are measured to be
\begin{eqnarray*}
B(\taupkz)  & = &  \brkoi, \\
B(\taupkzp) & = &  \brkoj, \\
B(\taukkz)  & = &  \brkok, 
\end{eqnarray*}
where the first error is statistical and the second systematic.
\end{abstract}

%\bigskip\bigskip
%\begin{center}
%Authors: R. Sobie and I. Lawson \\
%Editorial Board: R.Coxe, A.Stahl, M.Thomson, S. Towers, N. Wermes \\
%\vspace{1cm}
%{\Large \bf Final Draft, 22 October 1999 \\ }
%{\large \bf Draft version, do not quote \\}
%\vspace{0.8cm}
%{\bf Comments to: Randall.Sobie@cern.ch and Ian.Lawson@cern.ch \\ by
% 28 October 1999, 18:00 \\}
%\end{center}

\bigskip
\bigskip
\begin{center}{\large
%(Submitted to European Physical Journal C)
(To be submitted to European Physical Journal C)
}\end{center}
\end{titlepage}

\begin{center}{\Large        The OPAL Collaboration
}\end{center}\bigskip
\begin{center}{
%begin authorlist PLEASE DO NOT DELETE THIS COMMENT
G.\thinspace Abbiendi$^{  2}$,
K.\thinspace Ackerstaff$^{  8}$,
P.F.\thinspace Akesson$^{  3}$,
G.\thinspace Alexander$^{ 23}$,
J.\thinspace Allison$^{ 16}$,
K.J.\thinspace Anderson$^{  9}$,
S.\thinspace Arcelli$^{ 17}$,
S.\thinspace Asai$^{ 24}$,
S.F.\thinspace Ashby$^{  1}$,
D.\thinspace Axen$^{ 29}$
% ...
%end authorlist PLEASE DO NOT DELETE THIS COMMENT
}\end{center}\bigskip
\bigskip
%begin institutes
$^{  1}$School of Physics and Astronomy, University of Birmingham,
Birmingham B15 2TT, UK
\newline
$^{  2}$Dipartimento di Fisica dell' Universit\`a di Bologna and INFN,
I-40126 Bologna, Italy
\newline
$^{  3}$Physikalisches Institut, Universit\"at Bonn,
D-53115 Bonn, Germany
\newline
% ...
$^{  8}$CERN, European Organisation for Particle Physics,
CH-1211 Geneva 23, Switzerland
\newline
$^{  9}$Enrico Fermi Institute and Department of Physics,
University of Chicago, Chicago IL 60637, USA
\newline
% ...
$^{ 16}$Department of Physics, Schuster Laboratory, The University,
Manchester M13 9PL, UK
\newline
$^{ 17}$Department of Physics, University of Maryland,
College Park, MD 20742, USA
\newline
% ...
$^{ 23}$Department of Physics and Astronomy, Tel Aviv University,
Tel Aviv 69978, Israel
\newline
$^{ 24}$International Centre for Elementary Particle Physics and
Department of Physics, University of Tokyo, Tokyo 113-0033, and
Kobe University, Kobe 657-8501, Japan
\newline
% ...
$^{ 29}$University of British Columbia, Department of Physics,
Vancouver BC V6T 1Z1, Canada
\newline
% ...
%end notes
\bigskip

\newpage
\section{Introduction \label{sec:intro}}
\par

The large samples of $\PZz$
events collected at $\Pep\Pem$ colliders over the past ten years have
made it possible to study resonance dynamics and test low energy QCD using
the decays of $\tau$ leptons to kaons.
In this paper, measurements of the branching ratios of the
{\small $\taupkz$}, {\small $\taupkzp$} and {\small $\taukkz$} decay 
modes are presented.\footnote{Charge conjugation is implied throughout 
this paper.} These measurements are based on two samples that identify
$\tau$ decays with \PKzL\ and \PKzS\ mesons. The $\PKzL$ mesons are 
identified by their one-prong
nature accompanied by a large deposition of energy in the hadron calorimeter
while the $\PKzS$ mesons are identified through their decay into two charged
pions. The selected number of {\small $\taupkzkz$} decays is very small 
and is treated as background in this analysis. 

The results presented here are extracted from the
data collected between 1991 and 1995 at energies close to
the \PZz\ resonance, corresponding to an integrated luminosity of
163 pb$^{-1}$, with the OPAL detector at LEP. 
A description of the OPAL detector can be found in~\cite{opaldetector}.
The performance and particle identification capabilities of the
OPAL jet chamber are described in~\cite{opaltracker}.
The $\tau$ pair Monte Carlo samples used in this analysis are
generated using the KORALZ 4.0 package \cite{koralz}.
The dynamics of the $\tau$ decays are simulated with the 
Tauola 2.4 decay library \cite{tauola}.
The Monte Carlo events are then passed through the 
OPAL detector simulation \cite{gopal}.

%%%%%%%%%%%%%%%%%%%%%%%%%%%%%%%%%%%%%%%%%%%%%%%%%%%%%%%%%%%%%%%
\section{Selection of {\boldmath $\tt$} events \label{sec:tausel}}
%%%%%%%%%%%%%%%%%%%%%%%%%%%%%%%%%%%%%%%%%%%%%%%%%%%%%%%%%%%%%%%

The procedure used to select $\PZz \rightarrow \tt$ events is identical 
to that described in previous OPAL publications~\cite{opal:e,opal:tp}.
The decay of the $\mathrm{Z}^0$ produces two back-to-back taus.
The taus are highly relativistic so that the decay products
are strongly collimated. As a result it is convenient to
treat each $\tau$ decay as a jet,  where
charged tracks and clusters are assigned to cones of half-angle
$35^\circ$~\cite{opal:e,opal:tp}. In order to avoid regions of
non-uniform calorimeter response, the two $\tau$ jets are
restricted to the barrel region of the OPAL detector by requiring that the
average polar angle\footnote{In the 
          OPAL coordinate system the e$^-$ beam direction 
          defines the $+z$ axis, and the centre of the LEP ring defines
          the $+x$ axis.  
          The polar angle $\theta$ is measured from the $+z$ axis, 
          and the azimuthal angle $\phi$ is measured from the $+x$ axis.}  
of the two jets satisfies
$\overline{|cos\theta|} < 0.68$.
The level of contamination from 
multihadronic events ($\eeqq$) is significantly reduced by requiring not more 
than six tracks and ten electromagnetic clusters per event.
Bhabha $(\eeee)$ and muon pair $(\eemm)$ events are removed by
rejecting events where the total electromagnetic energy and the
scalar sum of the track momenta are close to the centre-of-mass energy.
Two-photon events, $\eeeeee$ or $\eeeemm$, are removed by 
rejecting events which have little visible energy in the electromagnetic
calorimeter and a large acollinearity angle\footnote{
  The acollinearity angle is the supplement of the angle between the
  decay products of each jet.} 
between the two jets.

A total of $\ntpr$  events are selected for the $\kl$ sample
and $\nta$ events are selected for the $\ks$ sample from the 
1991-1995 data set.
The different number of $\tt$ events in the two samples is due to 
different detector status requirements used in each selection.
The fraction of background from non-$\tau$ sources is 
$(1.6 \pm 0.1)$\%~\cite{opal:e,opal:tp}.

%%%%%%%%%%%%%%%%%%%%%%%%%%%%%%%%%%%%%%%%%%%%%%%%%%%%%%%%%%%%%%%
\section{Selection of $\taukl$ decays \label{sec:klong}}
%%%%%%%%%%%%%%%%%%%%%%%%%%%%%%%%%%%%%%%%%%%%%%%%%%%%%%%%%%%%%%%

The selection of $\tau$ decays into to a final state containing at
least one $\PKzL$ meson follows a simple cut-based procedure.
Some \PKzS\ mesons will be
selected in this \PKzL\ selection. 
In particular, \PKzS\ mesons that decay  
late in the jet chamber, the solenoid or the electromagnetic calorimeter, 
will be indistinguishable from \PKzL\ mesons and are 
considered to be part of the \PKzL\ signal.
The Monte Carlo simulation predicts that the $\PKz$ component 
identified by the $\PKzL$ selection is composed of 86\%
$\PKzL$ and 14\% $\PKzS$ mesons. 

\begin{figure}[!htb]
\vspace{-10mm}
\begin{center}
\mbox{\epsfig{file=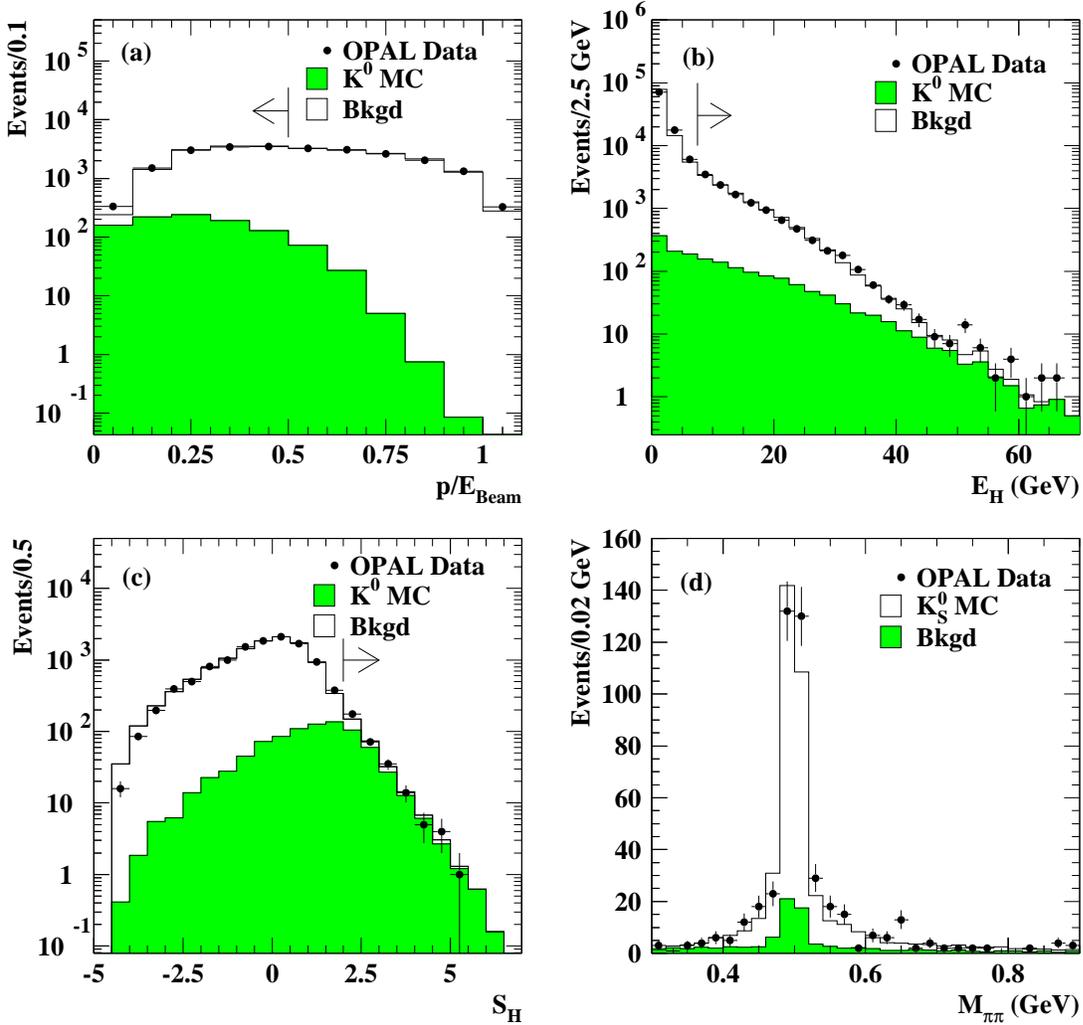,scale=0.69}}
\end{center}
\vspace{-10mm}
\caption{The first three plots show the $\PKzL$ selection variables:
(a) the momentum  divided by the beam energy $(p/\ebeam)$,
(b) the hadronic calorimeter energy ($\ehb$) and  (c)
$\shb = (E_{\mathrm{H}} - p)/\sigma_{\mathrm{H}}$. 
The fourth plot shows the mass distribution 
of jets which  pass the \PKzS\ selection.
In each case, jets which pass all of the selection
requirements except for the variable in question are plotted.
} 
\label{plot:precuts} 
\end{figure}

First, each jet must contain exactly one track pointing to the 
primary vertex and its momentum divided by the beam energy ($p/\ebeam$) 
must be less than 0.5, see fig.~\ref{plot:precuts}(a).
This requirement removes high-momentum pion decays from the one-prong sample.
In order to exclude leptonic background, there
must be at least one cluster in the hadron calorimeter and the total 
amount of energy measured by the hadronic calorimeter within the
jet must be greater than 7.5 GeV, 
see fig.~\ref{plot:precuts}(b).

The $\taukl$ decays will deposit on average more energy in the hadron
calorimeter than most other $\tau$ decays. 
This property is exploited using 
the variable, $\shb = (E_{\mathrm{H}} - p)/\sigma_{\mathrm{H}}$,
where $E_{\mathrm{H}}$ is the total energy deposited in the 
hadron calorimeter for the jet, $p$ is the momentum of the track 
and $\sigma_{\mathrm{H}}/E_{\mathrm{H}} = 0.165 + 0.847/\sqrt{E_{\mathrm{H}}}$ 
is the hadron calorimeter resolution. The energy is calibrated and
the resolution is measured  using pions from $\tau$ decays
that do not interact in the electromagnetic calorimeter.
Events with $\shb \ge 2.0$ are classified
as $\xm \; \klb$ decays, see fig.~\ref{plot:precuts}(c).
A total of $\nklong$  candidates are selected using the above requirements.
The background  is estimated to be 
\klinbg\%  from other $\tau$ decays and \klinqq\% from $\eeqq$ events.
The primary $\tau$ background consists of $\tau^- \rightarrow \pi^- \nu_\tau$, 
$\tau^- \rightarrow \rho(770)^- \nu_\tau$ and 
$\tau^- \rightarrow \mbox{a}_1(1260)^-\nu_\tau$ decays.

\begin{figure}
\begin{center}
\mbox{\epsfig{file=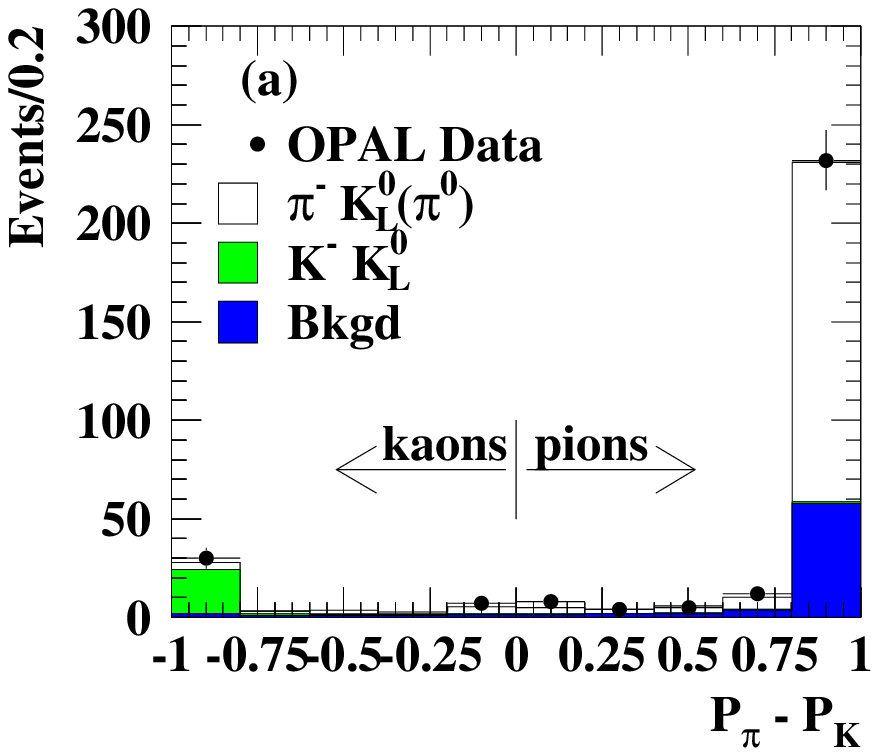,scale=0.75}}
\mbox{\epsfig{file=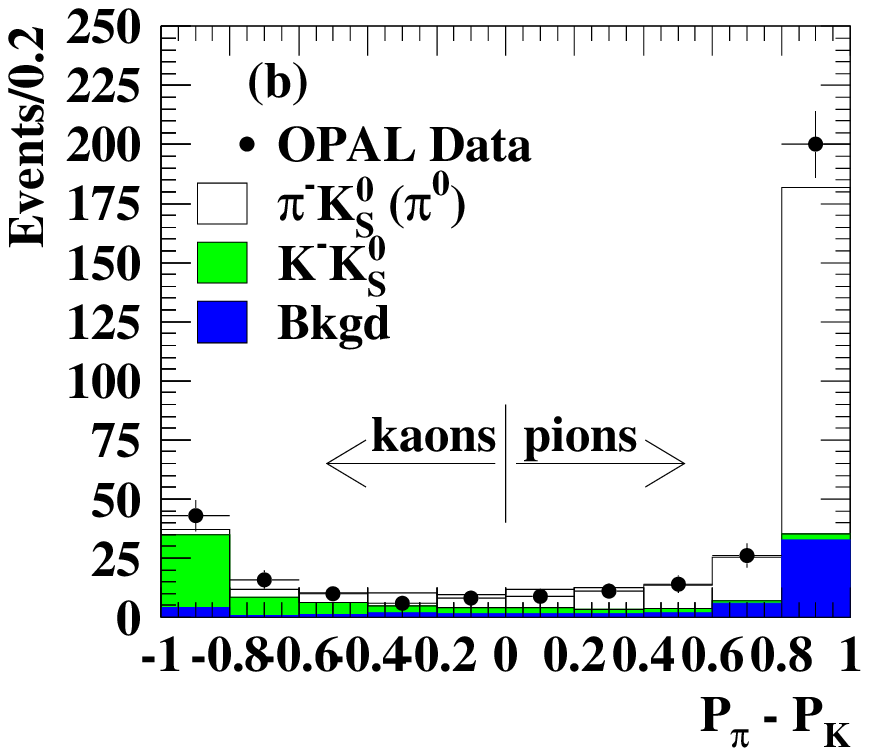,scale=0.75}}
\end{center}
\caption{
The $\pi/\PK$ separation variables. 
(a) shows the difference of the pion ($P_\pi$) and kaon 
($P_\PK$) probabilities, $P_\pi-P_\PK$, for the $\PKzL$ sample,
(b) shows  $P_\pi-P_\PK$ for the $\PKzS$ sample.}
\label{fig:dedx:kl}
\end{figure}

The sample of $\taukl$ decays is  subdivided into two sets:
one in which the track is identified as a pion and another in
which the track is identified as a kaon.
The sample with charged pions is then passed through an additional
selection which identifies those decays that include a $\pi^0$ meson.
The identification of the charged hadron uses the normalized
specific energy loss defined as 
$((\dedx)_{\mathrm{measured}} - (\dedx)_{\mathrm{expected}}) / \sigma_{\dedx}$,
where $\sigma_{\dedx}$ is the $\dedx$ resolution. Using this quantity, 
it is possible to separate charged pions and kaons 
at a level of $2\sigma$ in the momentum range of 2-30 GeV.
The expected $\dedx$ is calculated using the Bethe-Bloch equation 
parameterised for the OPAL jet chamber~\cite{opaltracker}. 
The parameterisation is checked using one-prong
hadronic $\tau$ decays by comparing the mean values and widths of the
normalised $\dedx$ distributions in bins of $\beta = p/E$, with $E^2 = p^2 + m_\pi^2$. 
It is found that a small $\beta$-dependent correction is to
be applied to the Monte Carlo. The correction shifts the mean value
of the expected $\dedx$ by up to 10\% and the widths by approximately
5\%. Charged pions and kaons are separated using a $\dedx$ probability
variable $W$, which is calculated from the normalized $\dedx$ for each
particle species. These are combined into pion and kaon probabilities,
\begin{eqnarray*}
\ppion & = & W_\pi/(W_\pi + W_\PK) \\
\pkaon & = & W_\PK/(W_\pi + W_\PK). 
\end{eqnarray*}
The distributions of the difference $\ppion-\pkaon$ is shown in 
fig.~\ref{fig:dedx:kl}(a).  A track
is considered to be a pion if $\ppion>\pkaon$.

\begin{figure}
\vspace{-5mm}
\begin{center}
\mbox{\epsfig{file=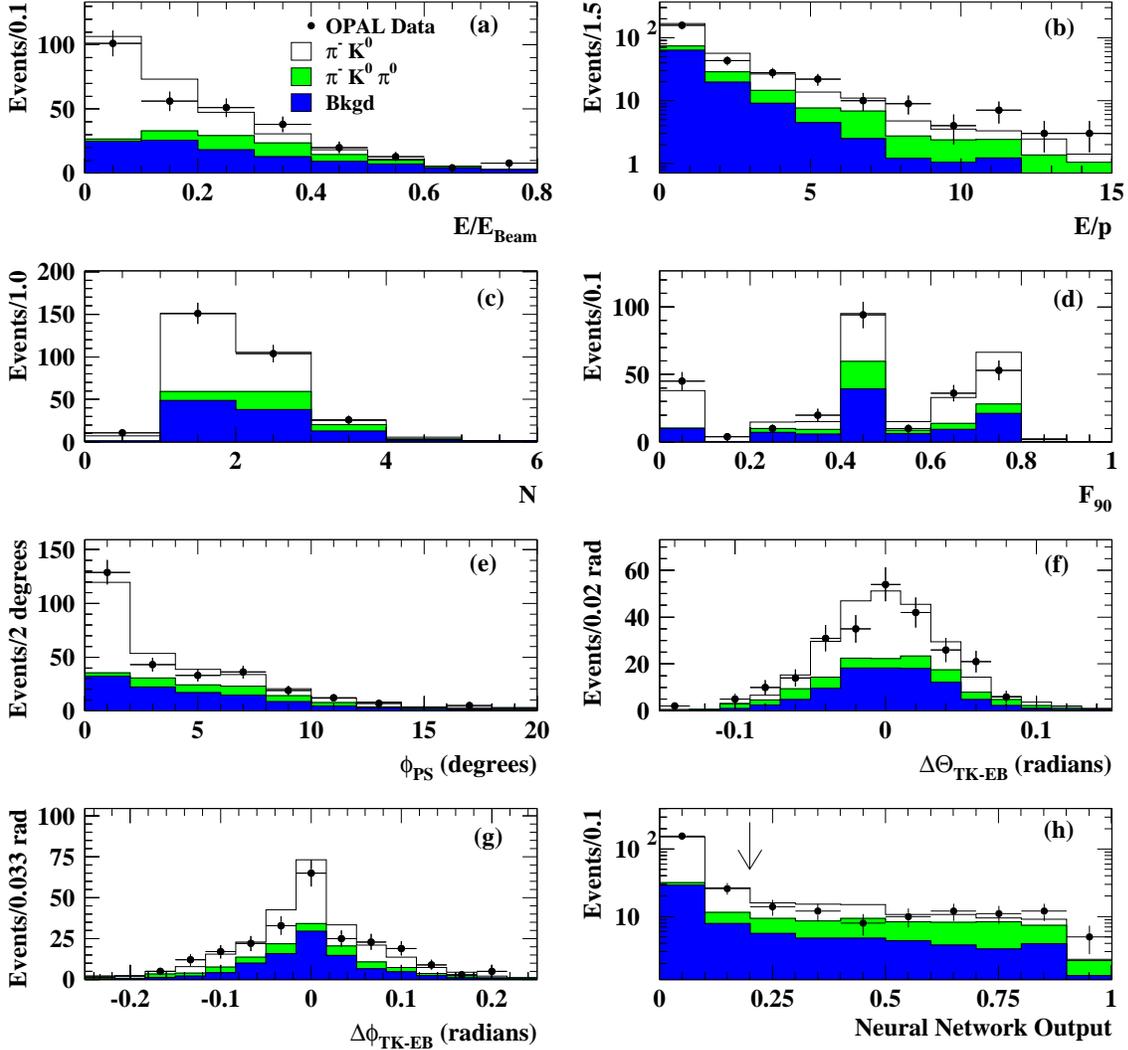,scale=0.65}}
\end{center}
\vspace{-5mm}
\caption{\label{plot:klnn}
The variables used in the neural network routine
for identifying $\pi^0$ mesons in the \PKzL\ sample:
(a) the electromagnetic energy divided by the beam energy;
(b) the ratio of the electromagnetic energy ($E$)
with the momentum of the track ($p$);
(c) the number of electromagnetic calorimeter clusters ($N$);
(d) the fraction of lead glass blocks in the electromagnetic calorimeter
with over 90\% of the energy in the jet; 
(e) angle between the position of the track at the presampler
and the presampler cluster furthest away from the jet axis; 
(f) and (g)  the difference in theta ($\Delta \theta$)
and phi ($\Delta \phi$) between the track and the vector obtained by adding all
the clusters in the electromagnetic calorimeter;
(h) the output of the neural network, the arrow indicates the cut
used to select decays containing $\pi^0$ mesons.
}
\end{figure}

A neural network algorithm is used to identify the $\taupkz$ and $\taupkzp$
decay modes.
The neural network algorithm uses 7 variables to identify the $\tau$ jets: 
\begin{itemize}

\item The total energy of the jet in the electromagnetic calorimeter divided
by the beam energy, $E/\ebeam$.

\item The total energy of the jet in the electromagnetic calorimeter divided
by the momentum of the track, $E/p$.

\item The number of electromagnetic clusters in the jet with an energy 
greater than 1 GeV.

\item The minimum fraction of active lead glass blocks which together
contains more than 90\% of the total electromagnetic energy of the jet,
$F_{90}$.

\item The difference in the azimuthal angle between the track and the 
presampler signal farthest away from the track but still within the jet,
$\phi_{\mathrm{PS}}$.

\item The difference in theta ($\Delta \theta$)
and phi ($\Delta \phi$) between the track and the vector obtained 
by adding together all the electromagnetic calorimeter clusters in the jet.

\end{itemize}
The variables used in the neural network and the output are
shown in fig.~\ref{plot:klnn}. If the neural network output is larger
than 0.2 then the decay is considered to contain a $\pi^0$ meson.

%%%%%%%%%%%%%%%%%%%%%%%%%%%%%%%%%%%%%%%%%%%%%%%%%%%%%%%%%%%%%%%
\section{Selection of $\tauks$ decays \label{sec:kshort}}
%%%%%%%%%%%%%%%%%%%%%%%%%%%%%%%%%%%%%%%%%%%%%%%%%%%%%%%%%%%%%%%

The algorithm for identifying $\ks$ candidates is similar to those
used in other OPAL analyses (for example, see \cite{opal:bose}).
The algorithm begins by pairing tracks with opposite charge.
Each track must have a minimum transverse momentum of 150 MeV and
more than 40 out of a possible 159 hits in the jet chamber.
Intersection points of track pairs in the plane perpendicular to the 
beam axis are considered to be secondary vertex candidates.
Each secondary vertex is then
required to satisfy the following criteria:
\begin{itemize}
\item 
   The radial distance $R_V$ from the secondary vertex to the
   primary vertex must be greater than 10 cm and less than 150 cm.

\item 
   The reconstructed momentum vector of the $\ks$
   candidate in the plane perpendicular to the beam axis must point to
   the beam axis within 1$^\circ$.

\item 
   If $R_V$ is between 30 and 150 cm
   (i.e. the secondary vertex is inside the jet chamber volume), then
   the radius of the first jet chamber hit associated with either of the two
   tracks ($R_1$) must satisfy $R_V - R_1 < 5$ cm.

\item 
   If $R_V$ is between 10 and 30 cm
   (not inside the jet chamber volume), then
   the impact parameter of the track is required to exceed 1 mm.

\item 
   The invariant mass of the pair of tracks, assuming both
   tracks to be electrons from a photon conversion, is required to be 
   greater than 100 MeV.
\end{itemize}

The $\tau$ jet is required to have at least one \PKzS\ candidate. If there
is more than one candidate then only the secondary vertex with an invariant 
mass closest to the true \PKzS\ mass is retained.
In addition, each jet is required to have only one additional
track, called the primary track.

A number of additional criteria are applied to reduce the
background from other $\tau$ decays.
Each track associated with the \PKzS\ 
must have $p>1$ GeV and must not have any hits in the axial regions of
the vertex drift chamber, which extends radially from 10.3 to 16.2 cm.  If
the radial distance to the secondary vertex is between 30 and 150 cm,
tracks with hits in the stereo region of the vertex drift chamber, which
extends radially from 18.8 to 21.3 cm, are rejected.

Candidate decays containing photon conversions identified with
an algorithm described in \cite{idncon}, are rejected.
Finally, the mass of the jet (assuming that the primary track is a pion)
must be less than 2 GeV and the invariant mass of the $\ks$ candidate 
is required to be between 0.4 and 0.6 GeV, see fig.~\ref{plot:precuts}(d).
A total of $\idt$  candidates are obtained with a 
background of approximately 10\%, consisting primarily of 
$\tau^- \rightarrow \rho(770)^- \nu_\tau$ and 
$\tau^- \rightarrow \mbox{a}_1(1260)^-\nu_\tau$ decays.

The sample of $\tauks$ decays is  subdivided into two sets:
one in which the primary track is identified as a pion and another in
which the primary track is identified as a kaon,
as described in section~\ref{sec:klong}.
In fig.~\ref{fig:dedx:kl}(b), the difference of the pion ($\ppion$) 
and kaon ($\pkaon$) 
probability weight ratios is shown for tracks identified as charged 
pions and kaons. A decay is considered to contain a $\pi^-$ if
$\ppion > \pkaon$.

\begin{figure}
\begin{center}
\mbox{\epsfig{file=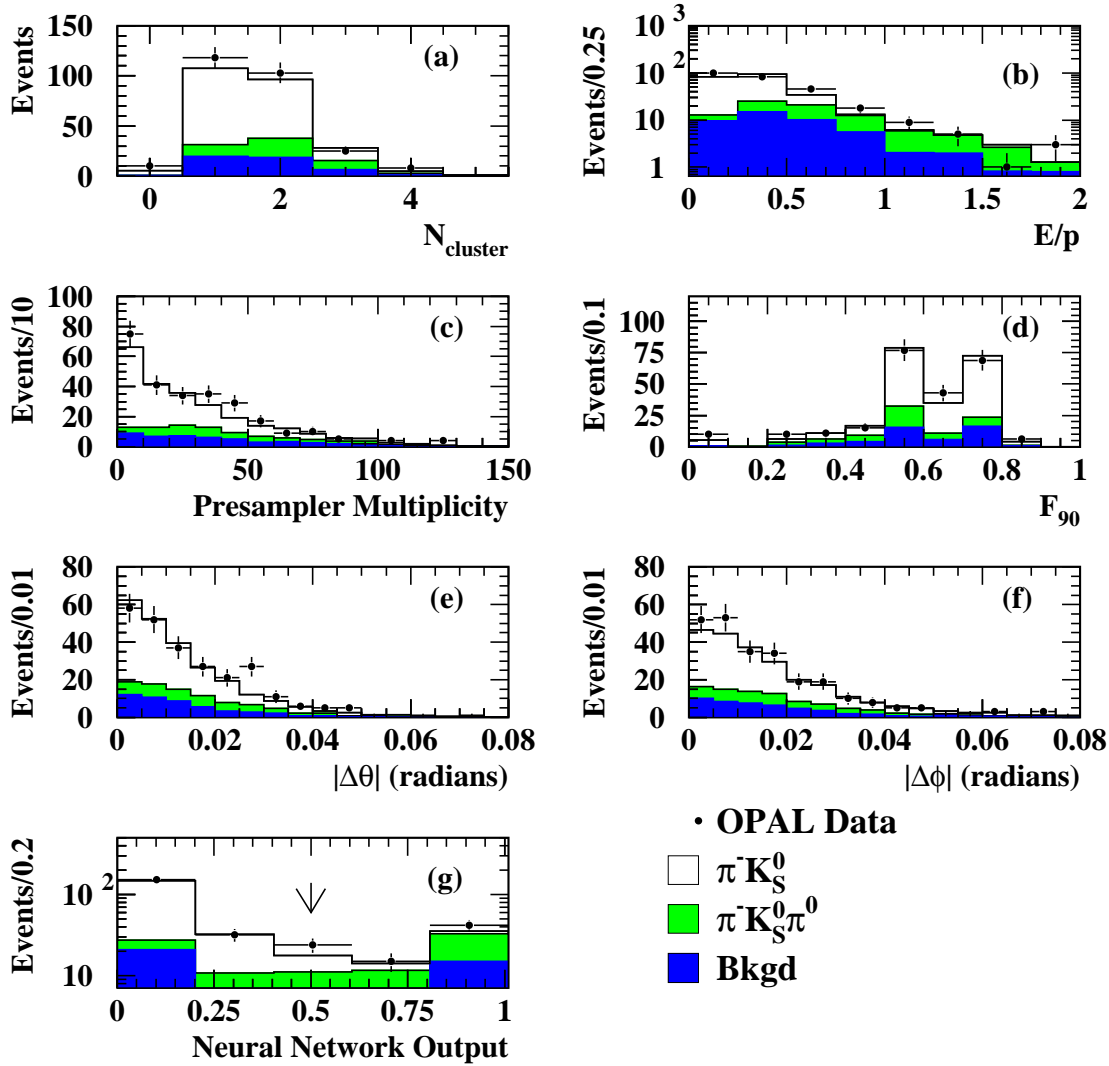,scale=0.80}}
\end{center}
\caption{\label{plot:nn}
The variables used in the neural network routine
for identifying $\pi^0$ mesons in the \PKzS\ sample:
(a) the number of clusters in the electromagnetic calorimeter;
(b) the ratio of the total energy in the electromagnetic calorimeter
divided by the total scalar momentum of the tracks;
(c) the presampler multiplicity,
(d) the fraction of lead glass blocks in the electromagnetic calorimeter
with over 90\% of the energy in the jet;  
(e) and (f)  the difference in theta ($\Delta \theta$)
and phi ($\Delta \phi$) between the vector obtained by adding
all the tracks and the vector obtained by adding all
the clusters in the electromagnetic calorimeter; 
(g) the output of the neural network, the arrow indicates the cut
used to select decays containing $\pi^0$ mesons.
}
\end{figure}

A neural network algorithm is used to identify the $\taupkz$ and $\taupkzp$
decay modes.
The neural network algorithm is similar to the one used in the $\PKzL$
analysis; 
the differences are due to the different topologies of the two
selections.
The neural network algorithm for this selection uses 6 variables to identify
the $\tau$ jets:
\begin{itemize}
\item The total energy of the jet in the electromagnetic calorimeter divided by 
the scalar sum of the momenta of the tracks, $E/p$.

\item The number of clusters in the electromagnetic calorimeter 
with energy greater than 1 GeV in the jet.

\item The minimum fraction of active lead glass blocks which together
contains more than 90\% of the total electromagnetic energy of the jet,
$F_{90}$.

\item The total presampler multiplicity in the jet.

\item The difference in theta ($\Delta \theta$)
and phi ($\Delta \phi$) between the vector obtained by adding
together all the tracks and the vector obtained by adding together all
the clusters in the electromagnetic calorimeter associated to the jet.

\end{itemize}
The variables used in the neural network algorithm are shown in 
figs.~\ref{plot:nn}(a-f).
A decay is considered to contain a $\pi^0$ if the neural 
network output, shown in fig.~\ref{plot:nn}(g), is greater than 0.5.

%%%%%%%%%%%%%%%%%%%%%%%%%%%%%%%%%%%%%%%%%%%%%%%%%%%%%%%%%%%%%%%
\section{Branching ratios \label{sec:br}}
%%%%%%%%%%%%%%%%%%%%%%%%%%%%%%%%%%%%%%%%%%%%%%%%%%%%%%%%%%%%%%%

The  branching ratios of the {\small $\taupkz$}, {\small $\taupkzp$} 
and {\small $\taukkz$} decay modes are calculated
independently for the $\tau$ jets containing \PKzL\ and \PKzS\ decays.
However, for a given data sample, the three branching ratios  are 
calculated simultaneously.
Each selection can be characterised in terms of the efficiency for 
detecting each decay mode $i$, the branching ratio of each mode and the number
of events selected in the data:
\[
\label{br_1}
\epsilon_{i1} B_1 + \epsilon_{i2} B_2 + \epsilon_{i3} B_3
+ \sum_{k=4}^{M} \epsilon_{ik} B_k =
\frac{N_i-N_i^{{\mbox{\scriptsize non}}-\tau}}{N_{\tau}
(1-f^{{\mbox{\scriptsize non}}-\tau})}
\]
where $N_i$ is the number of data events that pass the selection $i$,
$\epsilon_{ij}$ ($j=1,3$) are the efficiencies for selecting signal 
$j$ using selection $i$,
$\epsilon_{ik}$ ($k=4,\ldots,M$) are the efficiencies for selecting
the $\tau$ background modes using selection $i$ and $M$ is the number
of the $\tau$  decay modes.
The branching ratios of the signal channels and backgrounds are
$B_j$ ($j=1,3$) and $B_k$ ($k=4,\dots,M$), respectively.
The fraction of non-$\tau$ events in the $\tau$ pair sample is
$f^{{\mbox{\scriptsize non}}-\tau}$,
$N_\tau$ is the total number of taus in the data that pass the 
$\tau$ pair selection and $N_i^{{\mbox{\scriptsize non}}-\tau}$ is the
non-$\tau$ background present in each selection $i$.
The selection efficiencies ($\epsilon_{ij}$) for both signal and background
are determined from Monte Carlo simulation.
The $\tau$ background branching ratios are taken from the 
Particle Data Group compilation \cite{pdg}.

Solving the three simultaneous equations yields the 
branching ratios in each sample of selected events.
A small correction is applied to the branching ratios to account for 
any biases introduced into the $\tau$ pair sample by the $\tau$ pair selection.
The bias factor is defined as the ratio of the fraction of the selected decays in a
sample of $\tau$ decays after the $\tau$ selection is applied to the fraction
before the selection. The bias factors are calculated using approximately
2.2 million simulated $\tau^+\tau^-$ events.
The bias factors for the {\small $\taupkz$}, {\small $\taupkzp$} and 
{\small $\taukkz$} decays are found to be $\biexi$, $\biexj$ and $\biexk$
for the  branching ratios obtained from the $\kl$ sample, 
and  $\afbias$, $\bfbias$ and $\cfbias$
for the  branching ratios obtained from the $\ks$ sample. The
uncertainties on the measurements are statistical only.

The $\PKzL$ ($\PKzS$) selection identifies
$\nnkpi$ ($\adt$) {\small $\taupkz$} decays, $\nkpi0$ ($\bdt$) 
{\small $\taupkzp$} decays
and $\nkkpi0$ ($\cdt$) {\small $\taukkz$} decays.
The efficiency matrix for each sample is given in table~\ref{table:eff}.

The branching ratios obtained from the $\kl$ sample are
\begin{eqnarray*}
B(\taupkz)  & = &  \exbra , \\
B(\taupkzp) & = &  \exbrb , \\
B(\taukkz)  & = &  \exbrc ,\\[-3mm]
\end{eqnarray*}
while the branching ratios obtained from the $\ks$ sample are
\begin{eqnarray*}
B(\taupkz)  & = &  \abrz , \\
B(\taupkzp) & = &  \bbrz , \\
B(\taukkz)  & = &  \cbrz ,\\[-3mm]
\end{eqnarray*}
where the first error is statistical and the second is systematic.

%%%%%%%%%%%%%%%%%%%%%%%%%%%%%%%%%%%%%%%%%%%%%%%%%%%%%%%%%%%%%%%
\section{Systematic errors \label{sec:errors}}
%%%%%%%%%%%%%%%%%%%%%%%%%%%%%%%%%%%%%%%%%%%%%%%%%%%%%%%%%%%%%%%
The systematic uncertainties on the branching ratios are
presented in table~\ref{table:syst}. The dominant contributions
to the systematic uncertainty arises from the efficiency of
the two selections, the uncertainty of the backgrounds, the modelling
of the $\dedx$, the identification of the $\pi^0$ and the modelling
of Monte Carlo. These uncertainties are discussed in more detail 
below. In addition, 
there are straightforward contributions from the limited statistics
of the Monte Carlo samples used to estimate the selection efficiencies
and from the uncertainties on the bias factors. 
The systematic error on the branching ratios due to the Monte Carlo
statistics is calculated directly from the statistical uncertainties
on the elements of the inverse efficiency matrix~\cite{matrix:inversion}.
The systematic error on each branching ratio due to the
bias factor is calculated directly from the bias factor error.

\noindent{\bfseries \PKzL\ and \PKzS\ selection efficiencies:} 

\noindent
The \PKzL\ selection efficiency is sensitive to the calibration of the 
momentum, the energy measured by the hadron calorimeter and
the resolution of the hadron calorimeter.
The uncertainty on the momentum scale is typically better than
1\%~\cite{opal:tp}.
The uncertainty in the energy scale of the hadron calorimeter
is obtained by studying a sample of single charged hadrons from 
$\tau$ decays, the level of agreement between the data and Monte Carlo
is 1.5\%. The uncertainty due to the measurement of the resolution
of the hadron calorimeter is estimated by varying the resolution
within its uncertainties. Also, the shower containment is examined by looking at the
leakage of energy out of the back of the hadron calorimeter.
It is found that about 8\% of $\PKzL$ decays may not be
fully contained, these decays are well modelled by the Monte Carlo
and does not result in a systematic uncertainty.

The \PKzS\ selection efficiency is sensitive to the requirements on
the impact parameter, the momentum and the number of hits in the stereo 
and axial regions of the vertex chamber on the tracks associated to the \PKzS\ .
The systematic error on the \PKzS\ selection efficiency is determined
by dropping each relevant criterion except for the impact parameter resolution.
The impact parameter resolution has been shown to have an uncertainty that
is typically better than $\pm 20\%$~\cite{impact}. Variations of the
impact parameter resolution are found to have almost no contribution
to the systematic error on the \PKzS\ selection efficiency.

\noindent{\bfseries  Background estimation:} 

\noindent
The systematic error due to the background in the \PKzL\ sample
includes the uncertainty in the branching ratios of the background
decays, including the {\small $\taupkzkz$} 
and {\small $\taupkzkzp$} decays, as well as the uncertainty from the 
Monte Carlo statistics~\cite{pdg,aleph-k0}. 
The non-$\PKz$ background consists primarily of $\pi^-$, $\rho(770)^-$
and $\mathrm{a}_1(1260)^-$ decays in which the decays have a low momentum
track with at least one of the final $\pi$ mesons leaving some energy
in the hadron calorimeter. To investigate this background, the 
$\PKzL$ selection cut $S_H$  is reversed and the invariant mass spectra
are studied for each decay mode. The ratios of the data to the Monte Carlo
simulation are consistent with unity:
$0.97 \pm 0.02$, $1.04 \pm 0.02$ and $0.94\pm0.06$ for the 
$\pi^-\kzb$, $\pi^-\kzb \ge 1\pi^0$ and 
$\PKm\PKz\ge0\pi^0$ selections, respectively. The various contributions 
to the systematic error from the background are added in quadrature.

The background in the \PKzS\ sample includes {\small $\taupkzkz$} 
and {\small $\taupkzkzp$} decays,
which contain \PKzS\ mesons, and other $\tau$ decays, the uncertainty
is composed of the Monte Carlo statistical uncertainty plus a
component due to the uncertainty in the branching ratios of these 
decays~\cite{pdg,aleph-k0}.
A study of the sidebands of the $m_{\pi\pi}$
distribution (see fig.~\ref{plot:precuts}(d)) showed that the 
background prediction from other $\tau$ decays is  observed to be about 20\%
smaller in the Monte Carlo simulation than in the data. 
As a result, the background is scaled upward by a factor of 1.2
and a 20\% uncertainty is assigned to the background estimate.
The background estimate is cross-checked using the invariant mass 
distributions of the tracks associated with the \PKzS\ candidate for each 
of the exclusive channels. The ratios of the data to the Monte Carlo 
simulation are  consistent: $1.07 \pm 0.12$, 
$1.09 \pm 0.06$ and $0.93\pm0.11$ for the $\pi^-\ksb$, 
$\pi^-\ksb \ge 1\pi^0$ and $\PKm\PKzS\ge0\pi^0$ selections, 
respectively. The various contributions 
to the systematic error from the background are added in quadrature.

\noindent{\bfseries Modelling of $\dedx$:}  

\noindent
For both samples, the normalized $\dedx$ distributions are studied using the
sample of single charged hadrons from $\tau$ decays.
The uncertainty in the branching ratios is estimated by varying
the  means  of the normalised
$\dedx$ distributions by $\pm1$ standard deviation of their 
central values.  In addition, to account for possible differences
in the $\dedx$ resolution,
the widths of the normalized $\dedx$ distributions are varied by $\pm30$\%. 
Due to the three tracks present in the \PKzS\ sample, an additional
contribution to the systematic error is obtained by measuring the difference 
in the branching ratios when 
two different corrections are applied to the Monte Carlo. The first 
correction is
estimated from the one-prong hadronic tau decays while the second
correction is estimated using the sample of pions from the decay of the 
$\PKzS$. The various contributions 
to the systematic error from the $\dedx$ modelling are added in quadrature.

\noindent{\bfseries Identification of $\pi^0$:} 

\noindent
Both the \PKzL\ and \PKzS\ samples use a neural network algorithm
to separate the {\small $\taupkz$} and {\small $\taupkzp$} decay modes.
The most powerful variable for distinguishing between these two decays 
is the energy deposited in the electromagnetic calorimeter.
The systematic error in the branching ratio is evaluated by shifting 
the electromagnetic energy scale by $\pm 1\%$; this variation is assigned
after studying the differences between data and Monte Carlo in $E/p$ 
distributions for  3-prong $\tau$ decays. 

The uncertainty affecting the $\pi^0$ identification also includes the maximum
uncertainty  when each variable (except in those which
include the electromagnetic energy) is individually dropped 
from the neural network 
algorithm.  These uncertainties are added in quadrature with those obtained 
from the energy scale uncertainty. The stability of the neural network 
algorithm is studied by removing all but the two most significant
variables from the neural network, the results are within the systematic
uncertainties for both samples. As a cross check, the cut on the neural 
network output for both 
the \PKzL\ and \PKzS\ samples is varied between 0.1 and 0.8,
with the result being consistent within the total 
systematic uncertainties. 

\noindent{\bfseries Monte Carlo modelling:}

\noindent
The models used in the Monte Carlo generator can effect both the pion and
kaon momentum spectra. This effect can produce biases when
determining the $\PKz$ identification efficiency, the
$\PK/\pi$ separation and the $\pi^0$ identification.
The dynamics of the $\pi^-\kzb$ decay mode are
well understood. The $\pi^-\kzb$ decay mode is generated by 
Tauola via the $\PK^*(892)^-$ 
resonance. The $\PKm\PKz$ final state is generated by Tauola using phase
space only.

The $\taupkzp$ decay mode is composed of 
$\tm \! \rightarrow  \pi^- \kzb  \pi^0  \nut$ and
$\tm \! \rightarrow \! \pi^- \kzb \pi^0 \pi^0  \nut$ decays. 
The $\taupkp$ channel is modelled by Tauola assuming 
that the decay proceeds via the $\PK_1(1400)$ resonance. Recent results 
from ALEPH~\cite{aleph-oneprong} on one-prong $\tau$ decays with kaons,
and OPAL~\cite{sherry-towers} using 
$\tau^- \rightarrow \PKm \pi^-\pi^+\nu_\tau$ decays, suggest that
the $\taupkp$ decay will also proceed via the $\PK_1(1270)$ resonance.
A special Monte Carlo simulation is generated in which the final state
is created using the $\PK_1(1270)$ and $\PK_1(1400)$ resonances, using
the algorithm developed for the analysis described in~\cite{sherry-towers}.
The selection efficiency of the $\taupkp$ final state is estimated 
from the special Monte Carlo for both resonances. For the $\PKzL$ analysis,
the efficiencies  agree at a level of 10\%. For the $\PKzS$ analysis, the 
selection efficiencies agree at a level of 5\%.

The $\taupkpp$ decay mode is not modelled by Tauola. 
The branching ratio of this mode was recently measured to be
$(0.26\pm0.24)\times10^{-3}$~\cite{aleph-k0}. A special Monte Carlo
sample of the $\taupkpp$ decay mode is generated using flat phase space
and it is found  that the efficiency of the $\taupkpp$ decay mode agrees
within 30\% of the efficiency of the $\taupkp$ decay mode. For the 
systematic uncertainty associated with this decay mode, 30\% of the
$\taupkpp$ branching ratio is used.

The $\tm \rightarrow \, \km \kz {\mbox{\small $[\geq 1\pi^0]$}} \nut$ 
decay mode is composed of $\tm \rightarrow \, \km \kz \pi^0 \nut$ and
$\tm \rightarrow \, \km \kz \pi^0\pi^0 \nut$ decays.
The $\taukkc$ decay mode is generated by Tauola through a combination 
of the $\rho(1700)$ and $\mathrm{a}_1(1260)$ resonances.
Monte Carlo simulations of these two modes are generated
separately, again using the algorithm developed for the analysis described
in~\cite{sherry-towers}. The selection efficiencies of the $\taukkc$ decay mode 
are calculated for these two samples and are equivalent
within statistical errors. No systematic uncertainty is included for this 
channel. The $\taukkd$ decay mode  is not 
modelled by Tauola. The Particle Data Group~\cite{pdg} 
give an upper bound of $0.18\times10^{-3}$ for this channel.  A special Monte Carlo
sample of the $\taukkd$ decay mode is generated using flat phase space
and  the efficiency of the $\taukkd$ decay mode is observed to be within 30\% of
the efficiency of the $\taukkc$ decay mode. For the 
systematic uncertainty associated with this decay mode, 30\% of the $\taukkd$
branching ratio is used.

Finally, the $\taukkb$ selection efficiency may depend on the relative
$\taukka$ and $\taukkc$ branching ratios. Using the current world averages
from~\cite{pdg}, the relative contribution of each channel is varied
by $\pm 25\%$. For the $\PKzS$ analysis, no effect is observed
on the branching ratio, as the efficiency for selecting the two channels is
very similar; hence, no systematic error is included.

%%%%%%%%%%%%%%%%%%%%%%%%%%%%%%%%%%%%%%%%%%%%%%%%%%%%%%%%%%%%%%%
\section{Summary}
%%%%%%%%%%%%%%%%%%%%%%%%%%%%%%%%%%%%%%%%%%%%%%%%%%%%%%%%%%%%%%%

\begin{figure}
\vspace{-10mm}
\begin{center}
\mbox{\epsfig{file=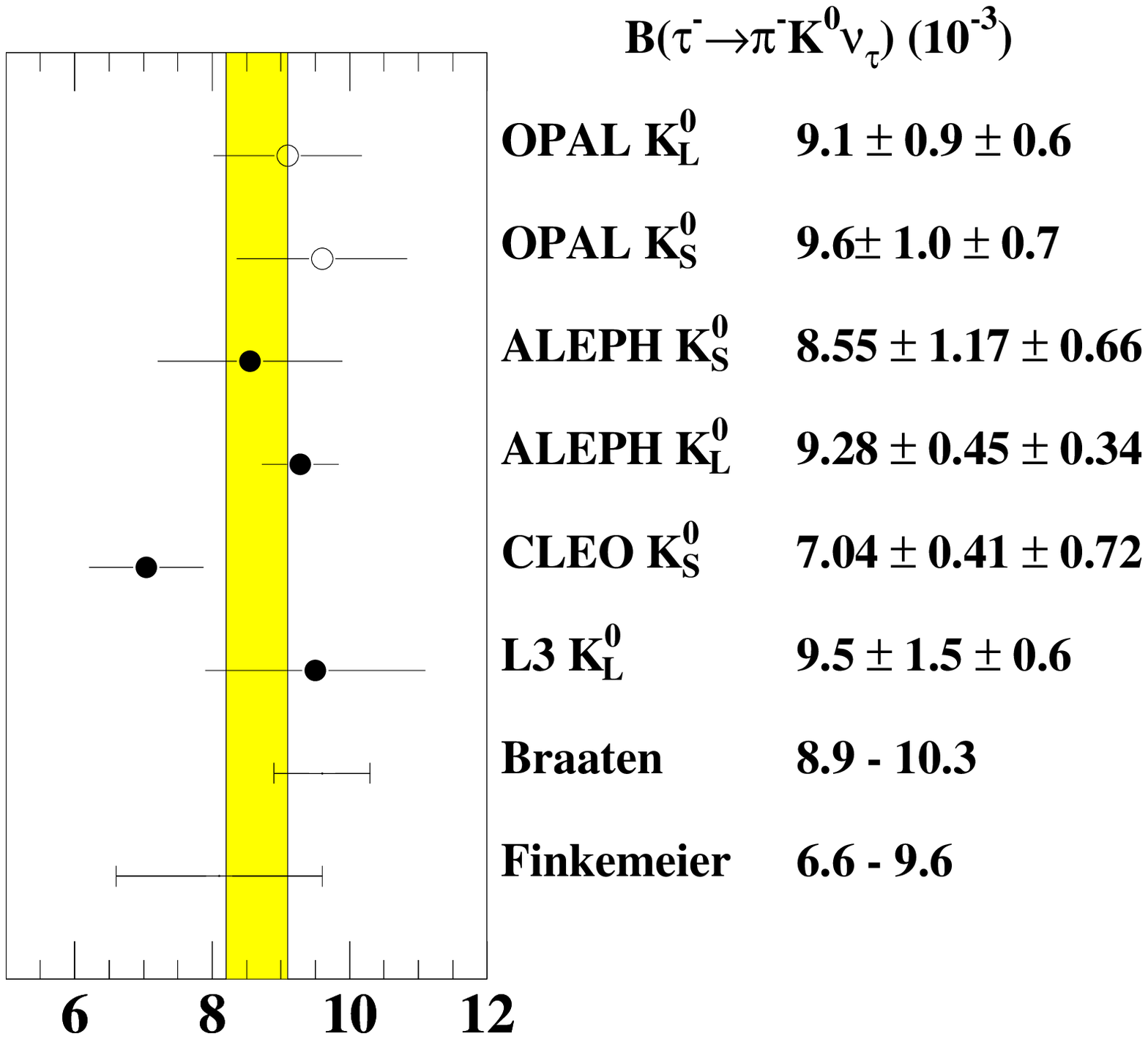,height=9.5cm}}
\mbox{\epsfig{file=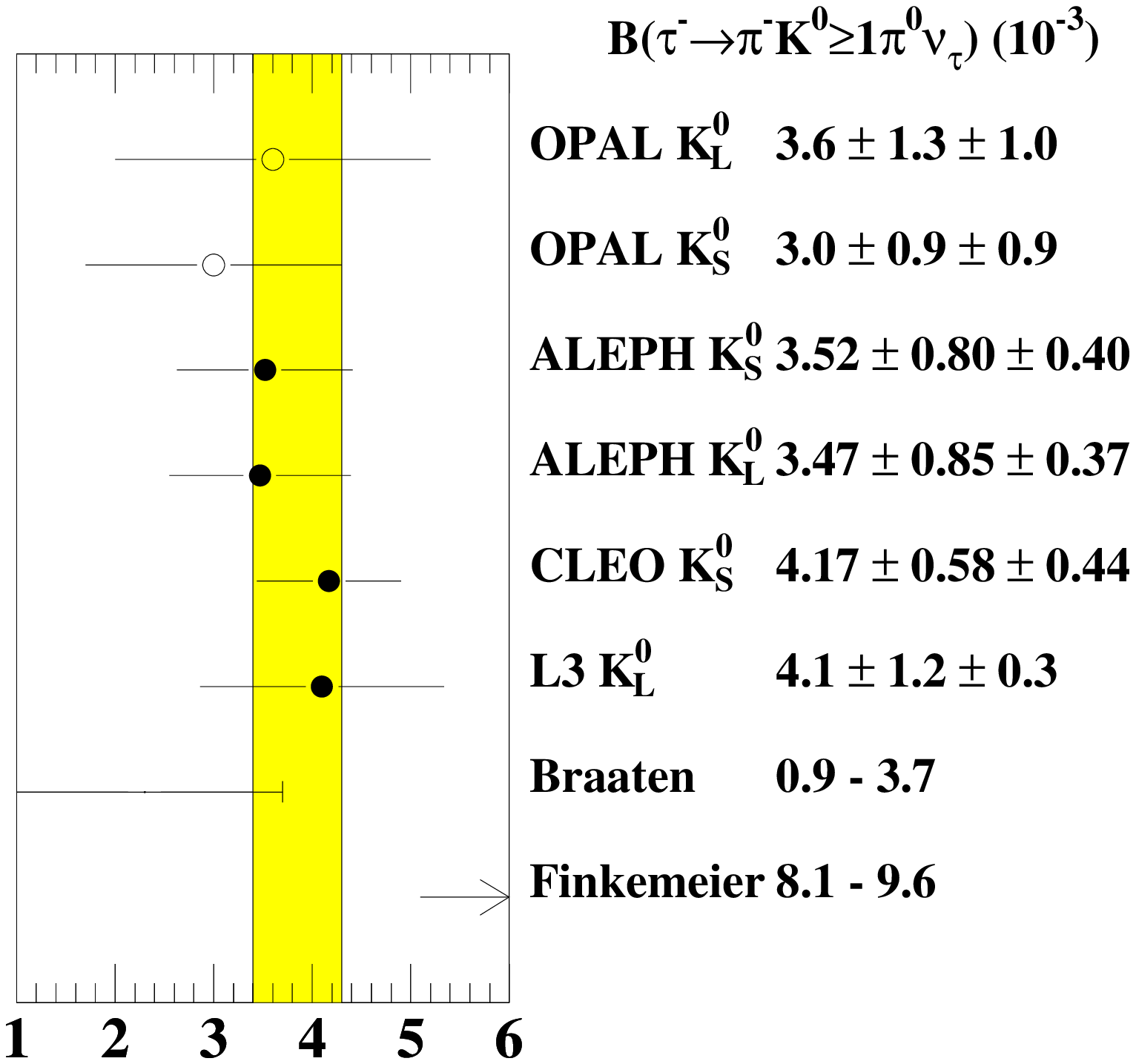,height=9.5cm}}
\end{center}
\caption{\label{plot:bratios}
Branching ratios of the  $\taupkz$ and $\taupkzp$ decays
measured or calculated to date. The solid band is the average branching ratio 
of the OPAL, ALEPH, CLEO and L3 measurements~\cite{aleph-oneprong,k0-measurents}.
The $\taupkzp$ results include both the $\taupkp$ and 
$\taupkpp$ measurements (ALEPH) and inclusive results (OPAL,L3,CLEO).
The theoretical estimates are shown for the $\taupkp$ decay mode 
only~\cite{langrangian,fink}.
The open points show the new OPAL results, the solid points other experimental
results and the bounded lines show two theoretical predicted
ranges of the branching fractions. }
\end{figure}

The branching ratios of the decays of the $\tau$ leptons to neutral kaons
are measured using the OPAL data recorded at centre-of-mass energies near
the $\mathrm{Z}^0$ resonance from a recorded luminosity of 163 pb$^{-1}$.
The measurement is based on two samples which identify $\tau$ decays
with $\PKzL$ and $\PKzS$ mesons.
The branching ratios obtained from the $\kl$ sample are 
\begin{eqnarray*}
B(\taupkz)  & = &  \exbra, \\
B(\taupkzp) & = &  \exbrb, \\
B(\taukkz)  & = &  \exbrc, 
\end{eqnarray*}
while the branching ratios obtained from the $\ks$ sample are 
\begin{eqnarray*}
B(\taupkz)  & = &  \abrz, \\
B(\taupkzp) & = &  \bbrz, \\
B(\taukkz)  & = &  \cbrz. 
\end{eqnarray*}
In each case the first error is statistical and the second systematic.
The combined results are
\begin{eqnarray*}
B(\taupkz)  & = &   \brkoi, \\
B(\taupkzp) & = &   \brkoj, \\
B(\taukkz)  & = &   \brkok. 
\end{eqnarray*}
The branching ratios are compared with existing measurements and
theoretical predictions in
fig.~\ref{plot:bratios} for the $\taupkz$ and $\taupkzp$ decay modes
\cite{aleph-oneprong,k0-measurents}.
The solid band is the new average branching ratio 
of the OPAL, ALEPH, CLEO and L3 measurements.
The results of this work are 
in good agreement with previous measurements. 

The branching ratios of the decay modes 
are predicted from various theoretical models. The measurement
of the decay fraction of the $\taupkz$ decay agrees well with the
range $(8.9-10.3) \times 10^{-3}$ estimated by Braaten \etal\ 
in~\cite{langrangian} and falls in the range of 
$(6.6-9.6) \times 10^{-3}$ predicted by Finkemeier and Mirkes in~\cite{fink}.
The decay $\taupkzp$, assuming that the decay contains only 
one $\pi^0$, is predicted to be in the range of $(0.9-3.7) \times 10^{-3}$ from 
\cite{langrangian} and in the range of $(8.1-9.6)\times 10^{-3}$ 
from~\cite{fink}. The $\taupkp$ branching ratio prediction by
Finkemeier and Mirkes is significantly higher than the
experimental results, however they argue that the widths of the
$\PK_1$ resonance~\cite{pdg} used in their calculation are unusually 
narrow and that increasing the $\PK_1$ width would give a prediction that
agrees with the experimental measurements~\cite{fink2}.

The branching ratio of the $\taukkz$ decay mode is 
the sum of the  $\taukka$ and $\taukkc$ decay modes. The decay fraction
agrees well with the estimated range $(2.4-4.0)\times10^{-3}$ predicted
by~\cite{langrangian} and $(2.3-2.7)\times10^{-3}$ predicted by~\cite{fink}.

\begin{figure}
\begin{center}
\includegraphics[scale=.75]{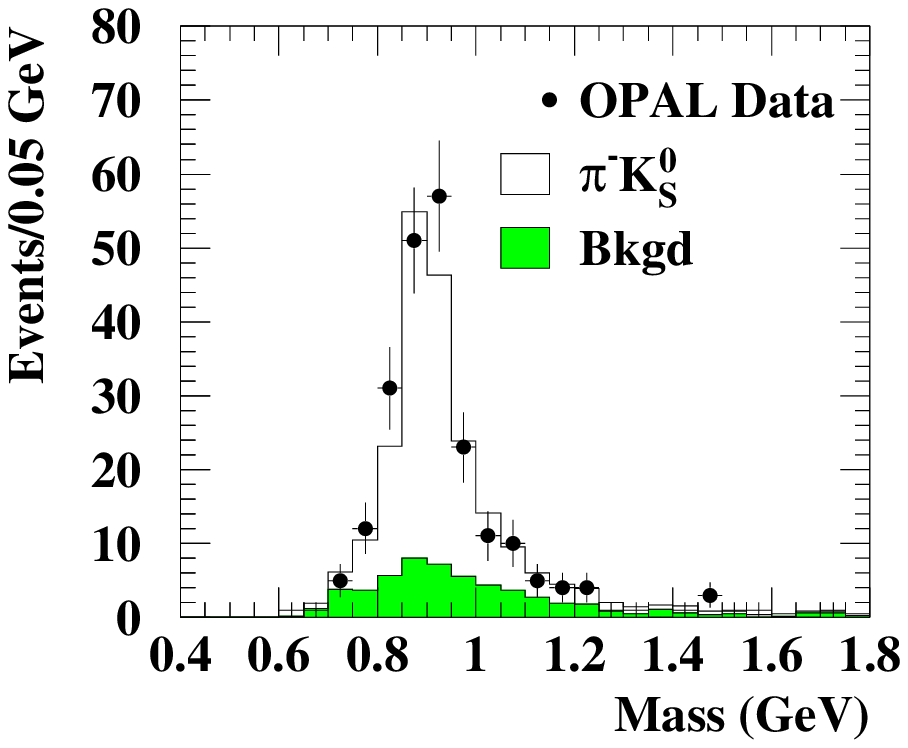}
\includegraphics[scale=.75]{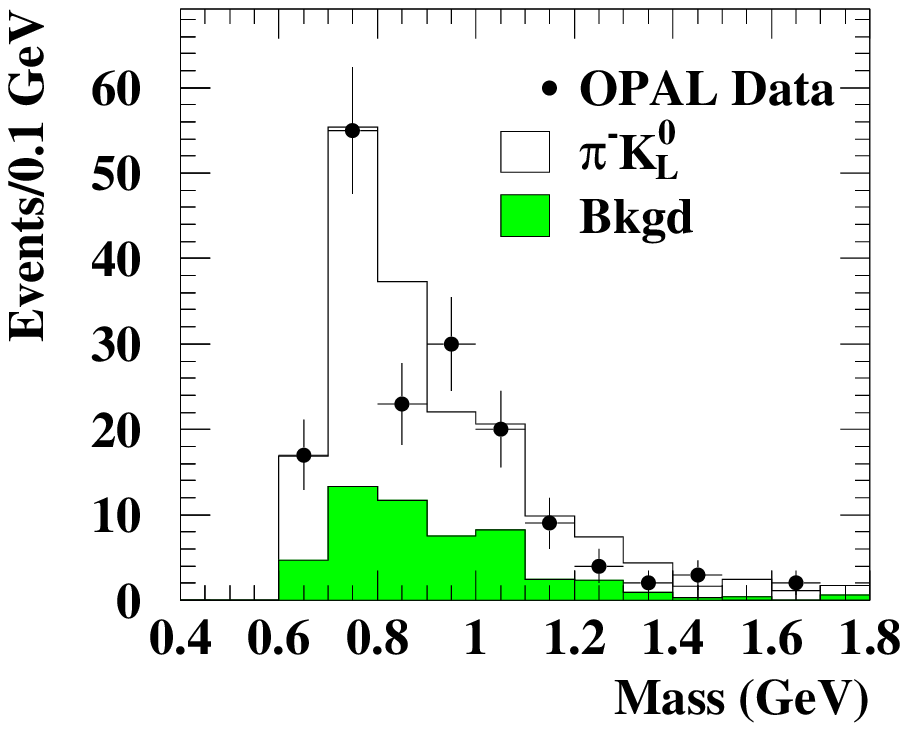}
\end{center}
\caption{\label{plot:kstmass}
The {\small $\ksb\pi^-$}  and {\small $\klb\pi^-$} 
invariant mass spectra for the the decay channels {\small $\taupks$} and 
{\small $\taupkl$}, respectively.
}
\end{figure}

The $\taupkz$ decay mode is assumed to be dominated by the $\PK^*(892)^-$ 
resonance. This can be observed from the $\pi^-\PaKz$ invariant 
mass distributions shown in fig.~\ref{plot:kstmass}, for the decay modes 
{\small $\taupks$} and {\small $\taupkl$}, respectively.
Assuming that the $\taupkz$ decay mode proceeds entirely through the
$\PK^*(892)^-$ resonance, then using isospin invariance
the branching ratio of the $\taukstar$ decay mode
is calculated to be $0.0140 \pm 0.0013$. 
This value is consistent with the current world average 
$0.0128 \pm 0.0008$~\cite{pdg}.

Finally, the ratio of the decay constants $f_\rho$ and $f_{\PK^*}$
can be estimated using the $\taukstar$ branching ratio and the OPAL
$\thpihz$  branching ratio of $0.2589 \pm 0.0034$~\cite{opal:tp}.
The $\thpihz$ decay mode is the sum of the decay
modes $\tau^- \rightarrow \pi^-\pi^0\nu_\tau$ and $\tau^- \rightarrow \PKm\pi^0\nu_\tau$.
The branching ratio of the $\tau$ to the final state $\PKm\pi^0\nu_\tau$
is calculated to be $(4.67 \pm 0.42)\times10^{-3}$ using isospin invariance 
and the $\taupkz$ branching ratio. Consequently, 
$B(\tau^- \rightarrow \pi^-\pi^0\nu_\tau)$ is derived to be $0.2543 \pm 0.0034$.  
Using these results, $\tan\theta_c = 0.227$ for the Cabibbo angle and the 
particle masses from~\cite{pdg}, the decay constant ratio
\begin{displaymath}
\frac{f_\rho}{f_{\PK^*}}  =  \tan\theta_c\sqrt{\frac{B(\trho)}{B(\taukstar)}}
\left(\frac{m^2_\tau - m^2_{\PK^*}}{m^2_\tau - m^2_\rho}\right)
\sqrt{\frac{m^2_\tau + 2 m^2_{\PK^*}}{m^2_\tau + 2m^2_\rho}} 
   =  0.93 \pm 0.05
\end{displaymath}
is obtained.
The error is dominated by the uncertainties on the branching
ratios. The recent result from ALEPH~\cite{aleph-k0}, 
$0.94 \pm 0.03$, agrees well with the new OPAL
result. Finally, this ratio has been predicted by Oneda~\cite{oneda} using
the Das-Mathur-Okubo sum rule relations~\cite{dmo} between the spectral
functions based on assumptions of $SU(3)_f$ symmetry. 
At the $SU(3)_f$ symmetry limit
($m_\mathrm{u} =m_\mathrm{d} =m_\mathrm{s}$), the decay constant ratio
is expected to be unity, $f_\rho = f_{\PK^*}$. In the asymptotic  $SU(3)_f$
symmetry limit at high energies, Oneda predicts that $f_\rho/f_{\PK^*}
=m_\rho/m_{\PK^*} = 0.86$.

\bigskip
%-----------------------------------------------------------------------
\section*{Acknowledgements}
%-----------------------------------------------------------------------
We particularly wish to thank the SL Division for the efficient operation
of the LEP accelerator at all energies
 and for their continuing close cooperation with
our experimental group.  We thank our colleagues from CEA, DAPNIA/SPP,
CE-Saclay for their efforts over the years on the time-of-flight and trigger
systems which we continue to use.  In addition to the support staff at our own
institutions we are pleased to acknowledge the  \\
Department of Energy, USA, \\
National Science Foundation, USA, \\
Particle Physics and Astronomy Research Council, UK, \\
Natural Sciences and Engineering Research Council, Canada, \\
Israel Science Foundation, administered by the Israel
Academy of Science and Humanities, \\
Minerva Gesellschaft, \\
Benoziyo Center for High Energy Physics,\\
Japanese Ministry of Education, Science and Culture (the
Monbusho) and a grant under the Monbusho International
Science Research Program,\\
Japanese Society for the Promotion of Science (JSPS),\\
German Israeli Bi-national Science Foundation (GIF), \\
Bundesministerium f\"ur Bildung, Wissenschaft,
Forschung und Technologie, Germany, \\
National Research Council of Canada, \\
Research Corporation, USA,\\
Hungarian Foundation for Scientific Research, OTKA T-029328, 
T023793 and OTKA F-023259.\\

%=======================================================================
%       References
%=======================================================================

%%%%%%%%%%%%%%%%%%%%%%%%%%%%%%%%%%%%%%%%%%%%%%%%%%%%%%%%%%%%%%%
% Tables
%%%%%%%%%%%%%%%%%%%%%%%%%%%%%%%%%%%%%%%%%%%%%%%%%%%%%%%%%%%%%%%
\newpage

%==============================================================
% Table:   Efficiency data
%==============================================================
\renewcommand{\arraystretch}{1.1}
\begin{table}
\begin{center}
\caption{\label{table:eff}
Signal efficiencies for each selection relative to selecting a $\PKz$. 
The errors on these efficiencies are based on Monte Carlo statistics only.
The first column lists the three selection classes. The three remaining
columns give the Monte Carlo selection efficiency for a decay of the
indicated type passing that selection classification.}
\vspace{0.5cm}
\begin{tabular}{l c c c} \hline
Selection & \multicolumn{3}{c}{Selection efficiency from MC (\%)}  \\ \hline
     & {\small $\taupkz$} 
     & {\small $\taupkzp$} 
     & {\small $\taukkz$}    \\ \hline
\multicolumn{4}{l}{$\kl$ sample} \\ 
$\pkz$    & $\exefii$ & $\exefij$ & $\exefik$  \\
$\pkzp$   & $\exefji$ & $\exefjj$ & $\exefjk$  \\ 
$\kkz$    & $\exefki$ & $\exefkj$ & $\exefkk$ \\ \hline
\multicolumn{4}{l}{$\ks$ sample} \\ 
$\pkz$    & $\effaa$  & $\effab$   & $\effac$   \\
$\pkzp$   & $\effba$  & $\effbb$   & $\effbc$   \\
$\kkz$    & $\effca$  & $\effcb$   & $\effcc$   \\  \hline
\end{tabular}
\end{center}
\end{table}
\renewcommand{\arraystretch}{1.0}

%==============================================================
% Table:   Systematic errors
%==============================================================
\renewcommand{\arraystretch}{1.1}
\begin{table}
\begin{center}
\caption{\label{table:syst}
Systematic errors on the branching ratios.}
\vspace{0.5cm}
\begin{tabular}{lccc} \hline
  & \multicolumn{3}{c}{Branching ratio systematic errors ($\times10^{-3}$) 
for the $\kl$ selection} \\ \hline
     & {\small $B(\taupkz)$} 
     & {\small $B(\taupkzp)$} 
     & {\small $B(\taukkz)$}    \\ \hline
MC statistics        & \exmcii\  & \exmcjj\   & \exmckk\ \\
Bias factor          & \exbiii   & \exbijj\   & \exbikk\ \\
$\kl$ efficiency     & \exklii\  & \exkljj\   & \exklkk\ \\
Background           & \exbtii\  & \exbtjj\   & \exbtkk\ \\
$\dedx$ modelling     & \exdeii\  & \exdejj\   & \exdekk\ \\
$\pi^0$ efficiency   & $\expOii$ & $\expOjj$  & 0.00     \\ 
MC modelling   & $\exmoii$ & $\exmojj$  & $\exmokk$ \\
\hline
Total                &  \exttii\ & \exttjj\   &  \exttkk\ \\ \hline
\multicolumn{4}{c}{} \\  
\multicolumn{4}{c}{} \\  \hline
 & \multicolumn{3}{c}{Branching ratio systematic errors ($\times10^{-3}$)  
for the $\ks$ selection} \\ \hline
     & {\small $B(\taupkz)$} 
     & {\small $B(\taupkzp)$} 
     & {\small $B(\taukkz)$}    \\ \hline
MC statistics        & $\axb$    & $\bxb$     & $\cxb$     \\
Bias factor          & $\axa$    & $\bxa$     & $\cxa$     \\
$\ks$ efficiency     & $\axd$    & $\bxd$     & $\cxd$     \\ 
Background           & $\axc$    & $\bxc$     & $\cxc$     \\ 
$\dedx$ modelling     & $\axe$    & $\bxe$     & $\cxe$     \\ 
$\pi^0$ efficiency   & $\axf$    & $\bxf$     & $\cxf$     \\
MC modelling   & $\axg$    & $\bxg$     & $\cxg$     \\
\hline
Total                & $\axt$    & $\bxt$     & $\cxt$     \\ \hline
\end{tabular}
\end{center}
\end{table}
\renewcommand{\arraystretch}{1.0}
%%%%%%%%%%%%%%%%%%%%%%%%%%%%%%%%%%%%%%%%%%%%%%%%%%%%%%%%%%%%%%%
% Figures
%%%%%%%%%%%%%%%%%%%%%%%%%%%%%%%%%%%%%%%%%%%%%%%%%%%%%%%%%%%%%%%

%==============================================================
% Figure: Klong x and E(HB)
%==============================================================
\setlength{\unitlength}{1mm}

%==============================================================
% Figure: Kl dE/dx
%==============================================================

%==============================================================
% Figure: Kl pizero selection variables
%==============================================================

%==============================================================
% Figure: Neural network variables 
%==============================================================

%==============================================================
% Figure: Branching ratios
%==============================================================
%==============================================================
% Figure: K* mass 
%==============================================================

%==============================================================
\end{document}